# Reflecting (on) the Nucleation Curve: Negative Surface Energy Stabilizes Nickel Ferrite Nano-Particles in Nuclear Reactor Coolant.

C.J. O'Brien, Zs. Rák and D.W. Brenner

Department of Materials Science and Engineering

North Carolina State University, Raleigh, NC 27695

Classical nucleation theory was first developed over eight decades ago (*1-3*), and with refinements and modifications since then (*4*) it remains the primary tool for understanding the kinetics of phase transformations. In its traditional form, classical homogeneous nucleation theory gives the change in free energy $\Delta G^T$ as a phase transformation proceeds as a sum of two terms. The first term is an interfacial energy per unit area $\gamma$ times the area between the two phases. The second term is the difference in free energy of the two phases $\Delta G$ times the volume of transformed material. Assuming that the transformation occurs through a spherical cluster of radius r and that $\gamma$ is isotropic, the free energy change as the transformation progresses is given by

$$\Delta G^T = 4\pi r^2 \gamma + \frac{4\pi}{3} r^3 \Delta G. \qquad (1)$$

A combination of positive surface energy and a negative $\Delta G$ yields a free energy barrier of $\frac{16\pi\gamma^3}{3(\Delta G)^2}$ at a critical radius $r^* = -\frac{2\gamma}{\Delta G}$ (Figure 1(a)). In general, the magnitude of $\Delta G$ increases as the temperature is lowered below that of the phase transition, which produces both a smaller nucleation barrier and critical cluster radius. For direction-dependent $\gamma$ values, a Wulff construction (*5*) can be used to determine appropriate volume and surface energies for a given cluster size.

Based on first principles calculations, Lodziana *et al.* recently suggested that the exothermic dissociative chemisorption of water with particular surfaces of θ-alumina can lead to negative surface energies (*6, 7*). Based on this result and supporting experimental evidence, they proposed that this negative surface energy may be responsible for a thermodynamic stabilization of porous alumina and may contribute to alumina's sintering resistance.

We recently used a thermodynamics-informed first principles (TIFP) scheme (*8, 9*) to calculate the temperature-dependent surface energies of nickel oxide NiO and nickel ferrite $NiFe_2O_4$, two compounds that are known to deposit on the fuel rods in nuclear pressurized water reactors (PWRs) (*10, 11*). As described in detail elsewhere (*9*), these calculations predict a negative surface free energy for nickel ferrite when formed from ions in solution under PWR conditions of temperature (~600K), pressure (155 bar), and species concentration. Under these conditions the thermodynamics of bulk nickel ferrite yields a positive change in free energy $\Delta G$ for formation of the solid from dissolved ions.

$$Ni^{2+} + 2Fe^{2+} + 4H_2O \xrightarrow{yields} NiFe_2O_4 + (6H^+)_{aq} + (H_2)_{aq} \quad (2)$$

Combining this with a negative surface energy changes the sign of the nucleation relation Eq. (1) so that the barrier in traditional nucleation theory becomes a well that thermodynamically stabilizes dissolved clusters (Figure 1(b)). We call this a reflected nucleation curve. While a size dependence of phase stability, including the influence of aqueous and humid conditions on surface energies (*12-15*) is well established, the influence of negative surface energies on cluster stability under conditions where bulk thermodynamics gives dissolution has not been previously recognized.

In the TIFP scheme effective chemical potentials (ECPs) $\mu^0(T)$ for the metals and oxygen are determined by solving a system of linear equations of the form

$$\Delta_f G^0_{A_xB_yO_z}(T,P) = E_{A_xB_yO_z}(0K) - \frac{z}{2}\mu_{O_2}(T) - x\mu_A^0(T) - y\mu_B^0(T), \quad (3)$$

where $\Delta_f G^0_{A_xB_yO_z}$ are experimental values of the Gibb's free energy of formation and $E_{A_xB_yO_z}$ are energies from Density Functional Theory (DFT) calculations at 0K.(*8*) The results presented here used ECPs that were determined from a least squares fit to data for NiO, ZnO, Fe$_2$O$_3$, Fe$_3$O$_4$, FeO(OH), Cr$_3$O$_4$, CoFe$_2$O$_4$, ZnFe$_2$O$_4$ and NiFe$_2$O$_4$.(*8*) ECPs for water and solvated metal cations are determined from the expressions

$$\left(\Delta_f G^0_{H_2O}(T,P)\right)_l = \left(\mu_{H_2O}(T,P)\right)_l - \frac{1}{2}\mu^0_{O_2}(T) - \mu^0_{H_2}(T) \quad (4)$$

and

$$\left(\Delta_f G^0_{M^{n+}}(T,P)\right)_{aq} = (\mu_{M^{n+}}(T,P))_{aq} - \mu_M^0(T) + \frac{n}{2}\mu^0_{H_2}(T) - n\left(\mu^0_{H^+}(T,P)\right)_{aq} \quad (5)$$

respectively, and the conventional chemical potential form

$$\mu_{H^+}(T) = \mu^0_{H^+}(T_r) + RT\ln(10^{-pH}) \quad (6)$$

is used for the solvated proton. This scheme avoids having to perform DFT calculations on H$_2$ and O$_2$ molecules, and provides a straight forward method for incorporating solvated phases into DFT calculations.

Surface energies were determined from DFT slab calculations that were carried out using the Vienna *Ab-initio* Simulation Package (*16-18*) and the generalized gradient approximation with the exchange-correlation functional of Perdew, Burke, and Ernzerhof (*19, 20*) plus on-site Coulomb interactions (GGA+*U*). The on-site Coulomb interactions were implemented using the formulation of Dudarev, *et al.* (*21*) in which the single parameter, $U_{eff} = U$-$J$, describes the Coulomb repulsion. Values of 4.5 eV and 6.0 eV were used for $U_{eff}$ for all Fe and Ni atoms, respectively, in the oxides (*12, 22*). Further details are given in references (*8*) and (*9*).

Plotted in Figure 2(a) is the energy of the nickel ferrite (111) surface as a function of temperature under PWR conditions of pressure = 155 bar, pH = 7.2, and concentrations $[Ni^{2+}]$ = 1.66×10$^{-14}$ and $[Fe^{2+}]$ = 4.17×10$^{-13}$ mol/kg (*23*). This was the lowest energy surface of the 36 surfaces studied, and through a Wulff construction it is predicted to be the only surface that would appear at equilibrium, consistent with experiment (*24*). Also plotted in Figure 2(b) is the change in free energy from Eq. (2) for the same conditions. Plotted in Figure 3 is the change in free energy of an octahedral cluster, calculated with the data in Figure 2, as a function of the characteristic length at different water temperatures. As discussed above, a negative surface energy and a positive change in bulk free energy from solvated ions to forming a solid cluster yields reflected nucleation curves that, instead of having free energy barriers, have free energy wells that stabilize formation of nickel ferrite clusters. Furthermore, the depth of the wells and size of the clusters associated with these wells vary significantly with temperature similar to the nucleation barriers and critical radii that generally become smaller the further the temperature is below the liquid-solid transition temperature.

This result has potentially important implications for measuring, understanding and controlling the contribution of nickel ferrite to the porous metal oxides that form on the fuel rod cladding in PWRs. The current understanding of the deposition process is that species are deposited from the coolant to the fuel cladding during subcooled nucleate boiling by micro-layer evaporation and dryout, a process by which evaporation into the vapor concentrates dissolved species (*25*). To reduce concentrations of various species from the coolant and hence mitigate this process, PWRs use a Chemical and Volume Control System (CVCS) that filters suspended particulates (which are thought to originate primarily from corrosion of the steam generator tubing) and removes dissolved ions with a mixed bed demineralizer (*26*). The coolant temperature and pressure in the CVCS is typically reduced from those in the reactor to avoid damage to the demineralizing resins and the water pumps. Our results suggest filtration strategies should consider not only relatively large suspended corrosion products, but also nanometer-scale nickel ferrite clusters that are predicted to be inherently stable within the coolant. The filtration should also take into account changes in stable cluster sizes that result from reductions in the coolant temperature in the CVCS. At still lower temperatures, the size and stability of the clusters both decrease significantly (c.f. Figure 3) such that they may not be readily observed in coolant after reactor cool down. Therefore coolant sampled during reactor operation but analyzed at lower temperatures may not show the same relatively large clusters present during reactor operation. Instead *in situ* measurements may be needed to observe these solvated clusters.

These calculations and their analysis in terms of classical nucleation theory have suggested a new fundamental relation – the reflected nucleation curve – that reveals a previously unrecognized aspect of the theory of phase stability. These calculations also suggest the presence of stable octahedral nickel ferrite clusters in PWR coolant that may not be observed in the coolant outside of service conditions, and that should be considered in designing strategies for purifying PWR coolant during reactor operation.

This research was supported by the Consortium for Advanced Simulation of Light Water Reactors (http://www.casl.gov), an Energy Innovation Hub (http://www.energy.gov/hubs) for Modeling and Simulation of Nuclear Reactors under U.S. Department of Energy Contract No. DE-AC05-00OR22725.

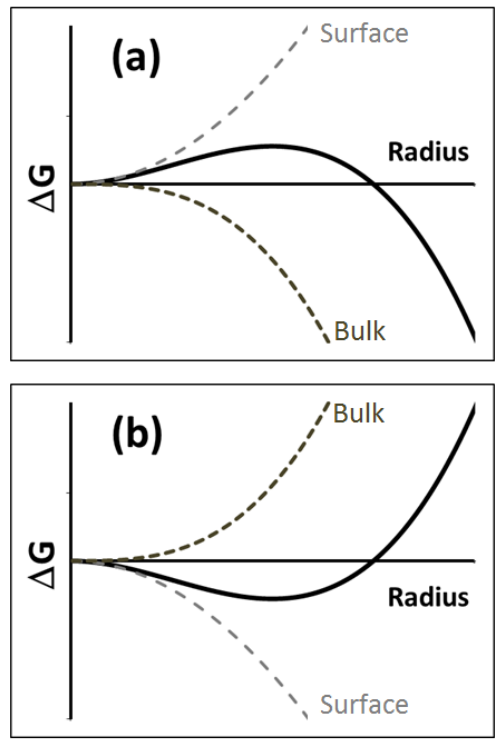

**Figure 1.** (a) Traditional homogenious nucleation curve with a positive surface energy and negative energy associated with the bulk leading to a nucleation barrier. (b) The reflected nucleation curve with a negative surface energy and positivebulk term.

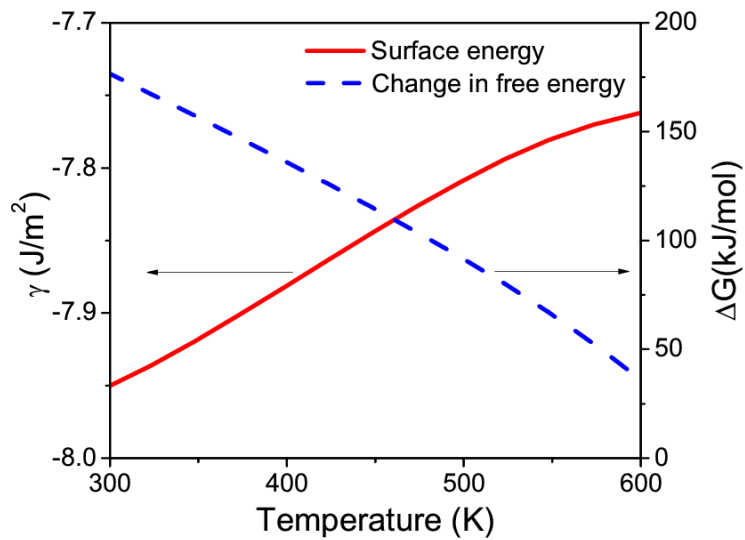

**Figure 2.** The energy of the (111) surface of NiFe$_2$O$_4$ as a function of temperature, under conditions of pressure, pH, and concentrations typical of PWR coolant (solid red line). The free energy of reaction for forming NiFe$_2$O$_4$, as described by Eq. (2) (dashed blue line).

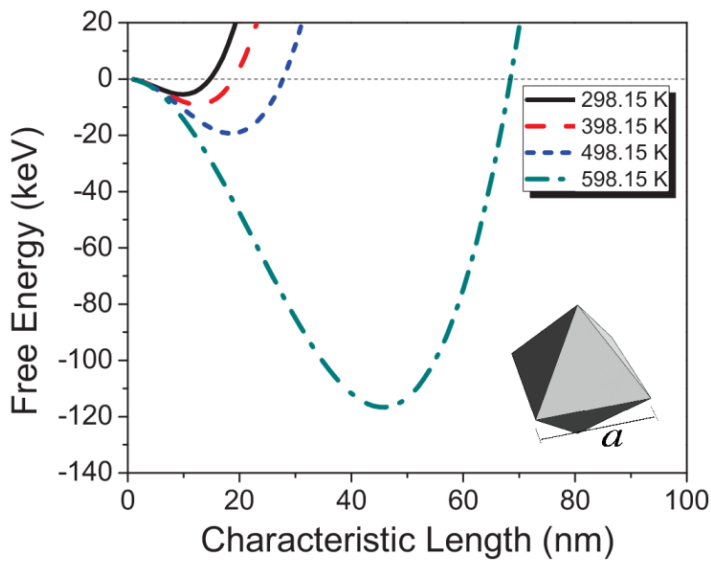

**Figure 3**. Change in free energy of an octahedral NiFe$_2$O$_4$ cluster as a function of its characteristic length, *a*, at different temperatures.

# References

1. R. Becker, W. Doring, *Ann Phys-Berlin* **24**, 719 (Dec, 1935).
2. M. Volmer, *Kinetik der Phasenbildung* (Theodore Steinkopff, Dresden, 1939).
3. J. B. Zeldovich, *Acta Physicochim Urs* **18**, 1 (1943).
4. V. I. Kalikmanov, *Nucleation Theory*. Lecture Notes in Physics (Springer, Dordrecht Heidelberg New York London, 2013), vol. 860.
5. G. Wulff, *Z Krystallogr Minera* **34**, 449 (Mar, 1901).
6. Z. Lodziana, N. Y. Topsoe, J. K. Norskov, *Nat Mater* **3**, 289 (May, 2004).
7. A. Mathur, P. Sharma, R. C. Cammarata, *Nat Mater* **4**, 186 (Mar, 2005).
8. C. J. O'Brien, Z. Rak, B. D. W., *Journal of Physics: Condensed Matter* **25**, 445008 (2013).
9. C. J. O'Brien, D. W. Brenner, *to be submitted*, (2013).
10. J. W. Yeon, I. K. Choi, K. K. Park, H. M. Kwon, K. Song, *J Nucl Mater* **404**, 160 (Sep 15, 2010).
11. J. A. Sawicki, *J Nucl Mater* **402**, 124 (Jul 31, 2010).
12. H. B. Guo, A. S. Barnard, *Phys Rev B* **83**, (Mar 8, 2011).
13. H. B. Guo, A. S. Barnard, *J Phys Chem C* **115**, 23023 (Nov 24, 2011).
14. H. B. Guo, A. S. Barnard, *J Mater Chem A* **1**, 27 (2013).
15. I. V. Chernyshova, S. Ponnurangam, P. Somasundaran, *Phys Chem Chem Phys* **15**, 6953 (2013).
16. G. Kresse, J. Furthmuller, *Comp Mater Sci* **6**, 15 (Jul, 1996).
17. G. Kresse, J. Furthmuller, *Phys Rev B* **54**, 11169 (Oct 15, 1996).
18. G. Kresse, J. Hafner, *Phys Rev B* **47**, 558 (Jan 1, 1993).
19. J. P. Perdew, K. Burke, M. Ernzerhof, *Phys Rev Lett* **77**, 3865 (Oct 28, 1996).
20. J. P. Perdew, K. Burke, M. Ernzerhof, *Phys Rev Lett* **78**, 1396 (Feb 17, 1997).
21. S. L. Dudarev, G. A. Botton, S. Y. Savrasov, C. J. Humphreys, A. P. Sutton, *Phys Rev B* **57**, 1505 (Jan 15, 1998).
22. A. Jain *et al.*, *Phys Rev B* **84**, (Jul 12, 2011).
23. J. Henshaw *et al.*, *J Nucl Mater* **353**, 1 (Jul 1, 2006).
24. Y. Cheng, Y. H. Zheng, Y. S. Wang, F. Bao, Y. Qin, *J Solid State Chem* **178**, 2394 (Jul, 2005).
25. H. Bindra, B. G. Jones, *Colloid Surface A* **397**, 85 (Mar 5, 2012).
26. R. Prince, *Radiation Protection at Light Water Reactors*. (Springer-Verlag, Berlin-Heidelberg, 2012).